\documentclass[10pt]{cai}

\usepackage{amsmath,amssymb,amsfonts}
\usepackage{graphicx}
\usepackage{textcomp}
\usepackage{xcolor}
\usepackage{amsmath,amsfonts,amssymb}
\usepackage{graphicx}
\usepackage{url}
\usepackage{hyperref}
\usepackage{times}
\usepackage{multirow}%
\usepackage{textcomp}%
\usepackage{listings}%
\usepackage{url}%
\usepackage{enumitem}%
\usepackage{textcomp}%
\usepackage{lipsum}
\usepackage{hhline}%
\usepackage{stfloats}%
\usepackage{etoolbox}%
\usepackage{soul}%
\usepackage[utf8]{inputenc}%
\usepackage{microtype}%
\usepackage{fancyvrb}%
\usepackage{subcaption}%
\usepackage{babel,blindtext}%
\usepackage{colortbl}%
\usepackage{mathrsfs}%
\usepackage{textcomp}%
\usepackage{manyfoot}%
\usepackage{booktabs}%
\usepackage{algpseudocode}%
\usepackage{xcolor}

\def\BibTeX{{\rm B\kern-.05em{\sc i\kern-.025em b}\kern-.08em
    T\kern-.1667em\lower.7ex\hbox{E}\kern-.125emX}}

\begin{document}
\begin{center}

\title{Enhancing Scientific Literature Chatbots with Retrieval-Augmented Generation: A Performance Evaluation of Vector and Graph-Based Systems}
\maketitle

\thispagestyle{empty}

\begin{tabular}{cc}
Hamideh Ghanadian\upstairs{\affilone,\affiltwo}, Amin Kamali\upstairs{\affilone,\affilthree}, Mohammad Hossein Tekieh\upstairs{\affilone,*}
\\[0.25ex]
{\small \upstairs{\affilone} Pest Management Regulatory Agency (PMRA), Health Canada, Ottawa, ON, Canada} \\
{\small \upstairs{\affiltwo} Faculty of Engineering, University of Ottawa, Ottawa, ON, Canada} \\
{\small \upstairs{\affilthree} Oral Health Branch (OHB), Health Canada, Ottawa, ON, Canada} \\
\end{tabular}
  
\emails{
  \upstairs{*}mohammad.tekieh@hc-sc.gc.ca
}
\vspace*{0.2in}

\end{center}

\begin{abstract}
This paper investigates the enhancement of scientific literature chatbots through retrieval-augmented generation (RAG), with a focus on evaluating vector- and graph-based retrieval systems. The proposed chatbot leverages both structured (graph) and unstructured (vector) databases to access scientific articles and gray literature, enabling efficient triage of sources according to research objectives. To systematically assess performance, we examine two use-case scenarios: retrieval from a single uploaded document and retrieval from a large-scale corpus. Benchmark test sets were generated using a GPT model, with selected outputs annotated for evaluation. The comparative analysis emphasizes retrieval accuracy and response relevance, providing insight into the strengths and limitations of each approach. The findings demonstrate the potential of hybrid RAG systems to improve accessibility to scientific knowledge and to support evidence-based decision making.
\end{abstract}



\section{Introduction}
Recent advances in artificial intelligence (AI) have led to the development of large language models (LLMs), which have revolutionized natural language processing (NLP) and conversational AI. LLMs, such as GPT-4 proposed by OpenAI \cite{radford2021chatgptplus}, are pre-trained on vast amounts of text data and can generate coherent, contextually relevant responses based on user queries. These models have been adopted across various domains, including scientific research, health care, and customer service, due to their ability to process and synthesize large volumes of information efficiently.

One of the most impactful applications of LLMs is in the development of chatbots. Chatbots leverage LLMs to facilitate human-like interactions, providing users with instant and informative responses. These systems have been used for customer support, education, and research assistance. In the scientific domain, chatbots equipped with retrieval-augmented generation (RAG) capabilities can significantly enhance literature reviews by retrieving and summarizing relevant research papers. 

Although LLMs and chatbots offer significant advantages in knowledge retrieval, they also face several challenges that can reduce their effectiveness. LLMs often generate responses based on probabilistic models, leading to possible misinformation or hallucinations. Ensuring that chatbot-generated responses are factually accurate and grounded in reliable sources remains a critical challenge. Traditional keyword-based search engines return large amounts of unstructured data, requiring users to manually filter relevant information. RAG-based chatbots need to efficiently retrieve the most relevant documents from large corpora to improve the quality of generated responses. It is essential to ensure that the retrieved documents align with user queries and that the generated responses accurately reflect the intended context. This involves optimizing retrieval strategies and refining generation mechanisms. The lack of standardized evaluation metrics for RAG-based systems makes it challenging to assess the effectiveness of different retrieval and generation approaches. Creating high-quality test sets and annotation frameworks is crucial for meaningful benchmarking.

To address these challenges, we propose an AI-powered literature chatbot that uses RAG to retrieve and synthesize responses from a database of scientific articles and gray literature. Our system facilitates paper triaging and knowledge extraction, enabling researchers to efficiently access relevant studies based on specific research queries. To evaluate the effectiveness of our system, we examine two primary use-case scenarios: retrieval from a single uploaded paper and retrieval from a larger, multi-paper database. Additionally, we compare the performance of different retrieval architectures: vector-based RAG, graph-based RAG, and a hybrid approach, to determine their efficacy in scientific literature retrieval.

Here are the main contributions of this article:
\begin{itemize}
\item We present a GPT-based implementation of multiple RAG systems and compare their retrieval and response-generation performance on targeted queries.

\item We generate two synthetic test datasets using a GPT model to evaluate and compare the RAG systems, covering use cases involving a single paper and a collection of papers.

\item We obtain subject-matter expert (SME) annotations for the test sets to ensure quality and apply performance metrics to assess the model's effectiveness in generating the test data.

\end{itemize}

\section{Background and Related Works}
In this section, we conduct a comprehensive review of the existing literature on the RAG system, encompassing research in the field of NLP. Additionally, we review the previous work that focused on evaluation of the RAG systems to provide a complete overview of their performance.
Retrieval-augmented generation (RAG) has become a popular approach in natural language processing, combining retrieval and generation models to improve knowledge-grounding, reduce hallucinations, and enhance personalization.

Lewis et al. \cite{lewis2020retrieval} introduced RAG to improve the performance of large language models in knowledge-intensive NLP tasks. Their proposed approach combines a pre-trained sequence-to-sequence (seq2seq) model (parametric memory) with a dense vector index of Wikipedia accessed via a neural retriever (non-parametric memory) for enhanced knowledge access and manipulation. The authors present two RAG formulations: one using the same retrieved passage for the entire generated sequence, and the other, allowing different passages per token. They demonstrate state-of-the-art results in several open-domain question-answering tasks, with improved performance on knowledge-intensive generation tasks. The key innovation lies in the end-to-end training of both the seq2seq generator and the retriever, enabling the system to learn to retrieve relevant information without explicit supervision.

Gao et al. \cite{gao2023retrieval} comprehensively reviewed RAG. The paper thoroughly details RAG's evolution through three paradigms: Naive RAG, Advanced RAG, and Modular RAG, analyzing the key components of retrieval, generation, and augmentation within each. Furthermore, it provides a current evaluation framework and benchmark, highlighting challenges and future research directions, including multimodal RAG and the interplay between RAG and LLM fine-tuning. The overall purpose is to provide a structured understanding of RAG's methodologies, applications, and future potential within the broader context of LLMs.

Jiang et al. \cite{jiang2024longrag} deployed RAG systems with long-context LLMs. LongRAG addresses the imbalance in traditional RAG systems by employing a "long retriever" to retrieve less than 8 highly relevant long (4K-token) units of information, rather than many short units. This approach significantly reduces the retriever's workload and minimizes the retrieval of irrelevant information. Then, a "long reader" uses these long units to generate answers via zero-shot inference with an existing long-context LLM. Experiments on four datasets, including Wikipedia-based and non-Wikipedia-based ones, demonstrated LongRAG's superior performance compared to traditional RAG methods and achieved state-of-the-art results without any fine-tuning. The study highlights the importance of reconsidering retrieval unit granularity in modern RAG systems to fully harness the capabilities of advanced LLMs.

Yu et al. \cite{yu2024evaluation} presented a survey of RAG systems, focusing on the challenges of evaluating these systems and proposing a unified framework for their evaluation. The paper emphasizes the complexity of RAG evaluation due to the hybrid structure of the retrieval and generation components and their dependence on dynamic knowledge sources.
The authors introduced a framework called the Auepora to address the challenges in RAG evaluation. They provided a comprehensive overview of RAG evaluation, outlining the challenges and introducing a structured approach for assessing these systems. By identifying gaps in current methodologies and suggesting future research directions, the authors aim to improve the effectiveness and user alignment of RAG system benchmarks.

Salemi et al. \cite{salemi2024evaluating} introduced eRAG, a novel method to evaluate retrieval models within RAG systems. Unlike computationally expensive end-to-end evaluation, eRAG assesses each retrieved document individually using the RAG system's LLM and a downstream task metric, yielding a higher correlation with overall RAG performance. Experiments across various datasets demonstrated eRAG's superior correlation and significant computational advantages over existing methods, showing improvements in Kendall’s correlation ranging from 0.168 to 0.494 and up to a 50-fold reduction in GPU memory usage.

\section{Methodology}
\begin{figure}[ht]
  \centering
  \ifpdf
    \includegraphics[width=\textwidth]{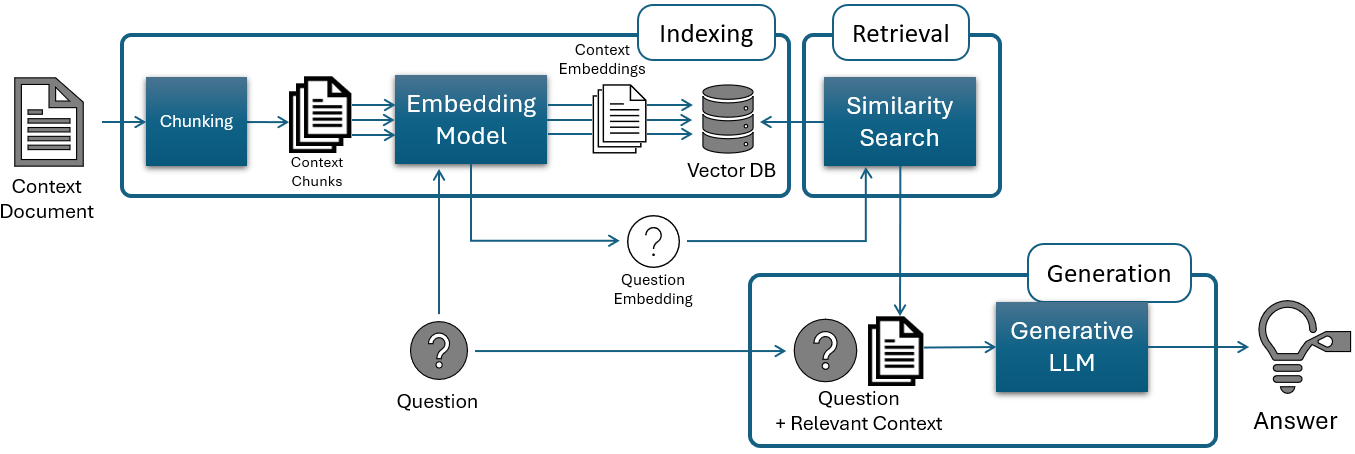}
  \else
    \includegraphics[scale=0.6,natwidth=330,natheight=120]{workflow.png}
  \fi
  \caption{Workflow of a Vector-based Retrieval-Augmented Generation}
  \label{fig: workflow}
\end{figure}

Retrieval-augmented generation (RAG) methodologies have demonstrated efficacy in incorporating current information, reducing inaccuracies (hallucinations) and improving the quality of generated responses, especially in specialized fields. In this section, we elaborate on each component of the RAG workflow. For each module, we determine the techniques incorporated into our pipeline. The workflow and corresponding methods for each module are illustrated in Figure \ref{fig: workflow}. Comprehensive experimental configurations, data sets, hyperparameters, and results are documented on our GitHub \footnote{\url{https://github.com/PMRA-DSA/dsa-public/tree/main/LitChat}} page.

In this study, we explore three retrieval approaches in the context of RAG systems: VectorRAG, GraphRAG, and a Hybrid RAG approach that combines vector-based and graph-based retrieval methods. Each method offers unique advantages based on its underlying retrieval techniques and data structures, enabling us to evaluate their performance across different tasks.

\textbf{VectorRAG} relies on vector-based retrieval, where document chunks are represented as high-dimensional embeddings. Using semantic search techniques, the system retrieves the most relevant chunks based on query-document similarity, and the model then generates a response by synthesizing information from the retrieved documents.

\textbf{GraphRAG} utilizes a property graph, retrieving relevant nodes and triplets from a graph database. The relationships between entities and concepts in the graph guide the model in generating a structured and logically coherent response, ideal for complex relationship-based queries.

\textbf{Hybrid RAG} is an advanced framework that combines both VectorRAG and GraphRAG to improve the accuracy and contextual relevance of information retrieval. The final output of the hybrid approach is a union of information from both VectorRAG and GraphRAG. 

These three methods form the basis of our experimental framework and we evaluate their performance in generating accurate and relevant responses using various test scenarios.

\subsection{Chunking}\label{sec: Chunking}
Chunking large data files into smaller manageable segments is essential to optimize LLM performance. This ensures that the model receives only the necessary information for the specific task, improving efficiency and relevance. This process can be applied at various levels of granularity, such as token, sentence, and semantic levels. Token-level chunking is a simple approach, but may break sentences, reducing retrieval quality. Sentence-level chunking breaks text into chunks by using multiple separators in a specified order. Semantic-level chunking leverages LLMs to determine natural segmentation points within text, maintaining contextual integrity, but demanding more computational resources. This study employs this advanced chunking technique to extract semantic relationships between text segments. The primary goal is to group portions of text with strong semantic ties, enhancing the cohesiveness of each segment. Unlike traditional methods that use fixed chunk sizes, the semantic splitter dynamically identifies optimal breakpoints based on embedding similarity. This adaptive approach ensures that each chunk consists of closely related sentences, resulting in a more meaningful and contextually rich segmentation \cite{fan2025trustrag}. In VectorRAG, we chose the default values for the semantic chunking parameters, which are a buffer size of 1 and a breakpoint percentile threshold of 95. In GraphRAG, the choice of text chunking method does not have a significant impact on the performance of a GraphRAG system because GraphRAG captures triplets from the text and does not directly rely on the text itself. However, the number of triplets extracted per chunk (maxPathsPerChunk) is critical for the model's performance. This number is directly related to the richness and depth of the resulting property graph \cite{chen2025pathrag}. Since extracting these triplets is a computationally expensive process (involving LLM calls), selecting an optimal number of triplets that balances model performance with extraction cost is essential. In the prototype, the default value for maxPathsPerChunk is set to 10.

\subsection{Embedding Model}\label{sec: Embedding Model}
Embeddings are numerical representations of text that capture semantic meaning by mapping words, phrases, or entire documents to high-dimensional vector spaces. These embeddings are crucial in NLP tasks, as they enable models to quantify semantic similarity and relationships between pieces of text. Selecting an appropriate embedding model is essential to ensure compatibility with the LLM used for downstream tasks. In this work, we used OpenAI’s  ``\textbf{\textit{text-embedding-ada-002}}'' model, known for its strong performance in generating high-quality embeddings that effectively capture semantic nuances, making it suitable for our semantic chunking approach \cite{oro2025comprehensive}.

\subsection{Database}\label{sec: Database}
Databases play a crucial role in managing and querying large data sets, offering diverse indexing and retrieval capabilities depending on the type of data and research requirements. For our research, we need a solution that could handle both embedding vectors and complex graph structures efficiently. We utilize ``\textbf{\textit{Neo4j}}'', a powerful graph database that supports vector and graph indexing. This choice allow us to seamlessly integrate semantic embeddings with graph-based data, enabling advanced querying and analysis through its dual indexing capabilities.

\subsection{Indexing Methods}\label{sec:indexing}

The indexing phase establishes the foundational data structures that enable efficient retrieval operations in both vector-based and graph-based RAG systems. This preprocessing step transforms raw documents into searchable representations that facilitate rapid query processing and contextual matching.\medskip

\textbf{Vector Index Construction:} Vector-based indexing converts textual documents into high-dimensional embedding representations using the embedding model. Each document is transformed into a dense vector that captures its semantic content, enabling similarity-based retrieval through proximity measures in the embedding space. These vectors are organized using efficient indexing structures such as approximate nearest neighbor algorithms to support scalable similarity search operations. In this work, we use the \textit{VectorStoreIndex} of \textit{LlamaIndex} to construct this index structure \cite{llamaindex_vectorstore}.\medskip

\textbf{Property Graph Construction:} Graph-based indexing extracts structured knowledge representations from unstructured text by identifying entities and their relationships. The resulting property graph consists of nodes representing entities and edges representing relationships, with each component containing associated metadata and properties. The metadata includes the text chunks from which the triplet was extracted. This structured representation enables traversal-based retrieval that can capture complex relational dependencies between concepts. In this work, we use \textit{PropertyGraphIndex} from \textit{LlamaIndex} to construct the property graph \cite{llamaindex_propertygraph}.

\subsection{Retrieval Methods}\label{sec:retrieval}

Retrieval methods identify and extract relevant documents or knowledge structures from the indexed corpus based on semantic similarity or structural relationships. These methods serve as the critical bridge between user queries and the generation process, determining which contextual information will inform the final response.\medskip

\textbf{Vector Index Retriever:} This method employs vector-based search to retrieve relevant information based on embedding similarity. By converting both queries and documents into high-dimensional vectors, the system measures semantic similarity through proximity metrics in the embedding space. The retriever selects the top-\textit{k} most relevant documents based on the closeness of their embedding vectors to the query vector, ensuring that semantically related content is prioritized for downstream processing. In this work, we rely on \textit{VectorIndexRetriever} from \textit{LlamaIndex} for this purpose \cite{llamaindex_vectorretriever}.\medskip

\textbf{Property Graph Retriever:} Graph-based retrieval leverages structured relationships within property graphs to identify relevant information. This method explores connections between entities by traversing graph edges, retrieving documents and data points that are relationally connected to the query concepts. The implementation combines \textit{VectorContextRetriever} \cite{llamaindex_vectorcontext}, which identifies nodes with high vector similarity to the query and traverses the graph to a specified depth parameter \textit{path\_depth}, with \textit{LLMSynonymRetriever} \cite{llamaindex_llmsynonym}, which performs keyword-based search using generated synonyms of query concepts to enhance coverage of relevant graph structures.\medskip

\textbf{Hybrid Retriever:} This approach integrates both vector-based and graph-based retrieval methodologies to leverage the complementary strengths of semantic similarity and structural relationships. By combining embedding-based proximity search with graph traversal mechanisms, the hybrid retriever provides comprehensive coverage of relevant information, capturing both semantically similar content and relationally connected knowledge.

\subsection{Generation Methods}\label{sec:generation}

The generation phase synthesizes retrieved contextual information into coherent, accurate responses that address the user's query. This process varies significantly between vector-based and graph-based approaches due to fundamental differences in the structure and representation of retrieved information.\medskip

\textbf{VectorRAG Generation:} In vector-based systems, the generation model processes the top-\textit{k} retrieved documents as concatenated textual contexts. The model synthesizes information from all selected documents simultaneously, combining semantic content from multiple sources to produce comprehensive responses. This approach enables the model to leverage the full breadth of retrieved textual information, generating answers that reflect the collective knowledge contained within the top-ranked documents.\medskip

\textbf{GraphRAG Generation:} Graph-based generation utilizes the structural relationships embedded within retrieved subgraphs and triplets. Rather than processing information as linear text, the generation model leverages the graph's relational structure to guide response synthesis. The model reasons about connections between entities and relationships, using the graph topology to maintain logical consistency and capture complex dependencies. This approach enables deeper analytical insights, particularly for queries involving multi-hop reasoning or complex relational patterns, as the model can trace logical pathways through the property graph structure to construct well-grounded responses.

\section{Datasets}
This section describes the datasets used in our research, which are divided into two main categories: the \textit{source documents} that populate the knowledge base and the \textit{test sets} used to evaluate the performance of our system in various scenarios.

\subsection{Source Documents}
The source documents form the primary knowledge base of our system. These documents are a collection of research articles obtained from scientific databases such as PubMed. These articles are studies on a variety of active ingredients registered for pesticide and herbicide products that cover various science areas, including ecotoxicology. We utilized web scraping and search engine APIs to gather documents from reputable sources, ensuring the quality and relevance of the information. The selection process involved defining relevant search queries, collecting documents and pre-processing them for indexing in the system. These documents are then converted into embeddings or stored as graph triplets, depending on the retrieval method used (VectorRAG or GraphRAG). This corpus serves as the foundation for generating responses to user queries in a RAG framework.

\subsection{Test Sets}
The test sets are designed to evaluate the performance and accuracy rate of the systems in two distinct test scenarios:\medskip

\textbf{RAG Applied to a Single Paper:} In this scenario, the system is tested on its ability to retrieve and generate responses using only the content of a single research article. This setup is useful for assessing the model's performance when the knowledge domain is highly specific and constrained to a single source. This dataset is generated synthetically using the \textit{GPT4o-mini} model and contains 500 sets of questions, answers, and contexts of selected papers. Research has shown that synthetic datasets can be a valuable resource for testing and data augmentation, offering a cost-effective and scalable approach to evaluating model performance \cite{ghanadian2024socially}.
This test set was annotated by a trained annotator to assess three key aspects:\medskip

\textit{Is the context related to the question?}

\textit{Is the answer derived from the context?}

\textit{Does the answer fully address the question?}\medskip

The annotator selected responses from \textit{Yes}, \textit{No}, or \textit{Unsure} for each question. After annotation, instances marked as \textit{No} or \textit{Unsure} were removed from the test set to ensure the quality and relevancy of the dataset.
Moreover, we evaluated the quality of the test set responses against the annotator's assessments to determine how effectively the \textit{GPT-4o-mini} model answered questions using the provided contexts. The detailed results are presented in Section \ref{sec: results}.\medskip

\textbf{RAG Applied to a Database:} This scenario evaluates the system's performance when retrieving information from a broader database containing multiple research articles and documents. This use case reflects a more typical application of RAG, where the model must navigate a larger corpus to find relevant information. Similarly to the previous test set, this dataset was generated synthetically using the \textit{GPT-4o-mini} model and includes 60 sets of questions, answers, and contexts derived from a wider collection of research papers. A subject matter expert assessed the data based on the testing criteria mentioned earlier. Responses were categorized as \textit{Yes}, \textit{No}, or \textit{Unsure}, and instances marked as \textit{No} or \textit{Unsure} were removed to maintain the quality of the dataset. The quality of the model’s responses was then evaluated against the expert's assessments. Detailed results are presented in Section \ref{sec: results}.

\section{Results}\label{sec: results}
This section presents the evaluation results of the proposed RAG systems, providing information on the performance of the systems.
\subsection{Test Set Evaluation}
In this part of the evaluation, we focused on assessing the quality of the questions and answers generated by the \textit{GPT-4o-mini} model based on the given context and related article.
The metric used in this evaluation is accuracy, which measures the proportion of correct answers produced by the model. Each answer is evaluated by an expert and labeled as either 1 (correct) or 0 (incorrect). Accuracy is calculated as the ratio of correctly generated answers to the total number of answers, providing a clear and straightforward measure of the model’s performance.
Table \ref{tab: test set} presents the performance evaluation of the synthetic datasets. Note that while the accuracy of the answers provided by the LLM to construct the synthetic data is reported here, we ultimately use the question-answer pairs approved by the subject matter expert.

\begin{table}[h]
\begin{tabular}{|c|c|c|}
\shortstack{\textbf{Test Dataset}\\{}} & 
\shortstack{\textbf{Is the Context related}\\\textbf{to the Question?}} &
\shortstack{\textbf{Does the Answer fully}\\\textbf{answer the Question?}} \\
\hline
 Single-Paper Test & 94.2\% & 92\% \\
\hline
 Multi-Paper Test & 81.6\% & 85\%
\end{tabular}
\vspace{0.2cm}
\caption{Evaluation of the quality of two synthetic test sets generated by LLM}
\label{tab: test set}
\end{table}

\subsection{Chatbot Evaluation}
This evaluation focuses on assessing the performance of the RAG in generating responses to each question based on the provided context. The evaluation examines how effectively the RAG utilizes the retrieved contexts to produce accurate, relevant, and complete answers. We utilized two metrics: cosine similarity and faithfulness, to measure the quality of the responses.
Cosine similarity evaluates how semantically similar the responses generated by the model are to the ground-truth answers.  Higher cosine similarity indicates a greater degree of semantic overlap between the two.
Faithfulness assesses whether the generated responses are grounded in the provided context, ensuring that the answers accurately reflect and rely on the retrieved information \cite{malin2024review}. This metric helps identify hallucinations or content that is not supported by the source material. A higher faithfulness score indicates that the responses are more aligned with and dependent on the retrieved evidence.
Together, these metrics provide a comprehensive evaluation of the RAG system's ability to deliver responses that are not only factually accurate but also contextually grounded and semantically consistent.
The results displayed in Table \ref{tab: RAG Eval} present a detailed analysis of the RAG system's response quality using the \textit{GPT-4o-mini} model with vector, graph, and hybrid RAG settings.

\begin{table*}[]
\begin{tabular}{c|cccc|cccc}
\multicolumn{1}{l|}{} &
  \multicolumn{4}{c|}{\textbf{RAG on Single Paper}} &
  \multicolumn{4}{c}{\textbf{RAG on Database}} \\ \cline{2-9} 
\textbf{Methods/Metrics} &
  \multicolumn{2}{c|}{Cosine Similarity} &
  \multicolumn{2}{c|}{Faithfulness} &
  \multicolumn{2}{c|}{Cosine Similarity} &
  \multicolumn{2}{c}{Faithfulness} \\ \cline{2-9} 
 &
  \multicolumn{1}{c|}{Mean} &
  \multicolumn{1}{c|}{SD.} &
  \multicolumn{1}{c|}{Mean} &
  SD. &
  \multicolumn{1}{c|}{Mean} &
  \multicolumn{1}{c|}{SD.} &
  \multicolumn{1}{c|}{Mean} &
  SD. \\ \hline
VectorRAG &
  \multicolumn{1}{c|}{0.670} &
  \multicolumn{1}{c|}{0.189} &
  \multicolumn{1}{c|}{0.841} &
  0.137 &
  \multicolumn{1}{c|}{0.651} &
  \multicolumn{1}{c|}{0.188} &
  \multicolumn{1}{c|}{0.830} &
  0.133 \\ \hline
GraphRAG &
  \multicolumn{1}{c|}{0.654} &
  \multicolumn{1}{c|}{0.184} &
  \multicolumn{1}{c|}{0.785} &
  0.153 &
  \multicolumn{1}{c|}{0.633} &
  \multicolumn{1}{c|}{0.174} &
  \multicolumn{1}{c|}{0.761} &
  0.149 \\ \hline
Hybrid RAG &
  \multicolumn{1}{c|}{0.687} &
  \multicolumn{1}{c|}{0.187} &
  \multicolumn{1}{c|}{0.845} &
  0.119 &
  \multicolumn{1}{c|}{0.662} &
  \multicolumn{1}{c|}{0.196} &
  \multicolumn{1}{c|}{0.834} &
  0.121 \\ \hline
\end{tabular}
\vspace{0.2cm}
\caption{Evaluation of the RAG performance on the two synthetic test datasets}
\label{tab: RAG Eval}
\end{table*}

\section{Discussion}
This study evaluates the performance of RAG systems using different retrieval approaches, specifically Vector-based RAG, Graph-based RAG, and a Hybrid RAG that combines both methods. The AI-powered literature chatbot was tested in two use-case scenarios: retrieval from a single uploaded paper and retrieval from an extensive database. The evaluation focused on the response faithfulness and the cosine similarity. Our results demonstrate the effectiveness of RAG approaches in improving access to scientific knowledge and supporting evidence-based decision-making.

The quality of the synthetic test sets was assessed to ensure an accurate benchmarking of RAG performance. The single-paper test set showed strong contextual relevance, with 94.2\% of the contexts being related to the question and 92\% of the answers fully addressing the question. The multi-paper test set had slightly lower scores, with 81.6\% context relevance and 85\% completeness in the answers. These differences suggest that while single-paper retrieval is relatively straightforward, multi-document retrieval presents additional challenges, such as increased ambiguity and the need for more sophisticated disambiguation techniques.

The results show that the three retrieval methods - VectorRAG, GraphRAG, and Hybrid RAG - performed well in retrieving relevant documents. However, Hybrid RAG consistently outperformed both individual methods in terms of retrieval accuracy and response faithfulness. This suggests that combining vector-based and graph-based retrieval techniques enhances the system's ability to retrieve contextually relevant information while maintaining logical coherence.

In the chatbot evaluation, responses were assessed based on the cosine similarity and faithfulness to the retrieved context. The Hybrid RAG approach achieved the highest scores in both metrics, with a mean cosine similarity of 0.687 and faithfulness of 0.845, indicating that it generated more contextually accurate responses. VectorRAG performed slightly better than GraphRAG in cosine similarity (0.670 vs. 0.654) and faithfulness (0.841 vs. 0.785), probably due to the direct semantic matching used in vector-based retrieval. GraphRAG, despite scoring lower, provided more structured and logically coherent responses, making it particularly useful for complex queries involving relationships between concepts.

These findings highlight several key insights:

\begin{itemize}
    \item \textbf{Hybrid retrieval improves answer accuracy:} By leveraging both semantic similarity and structured knowledge representations, Hybrid RAG maximizes retrieval efficiency and relevance.
    \item \textbf{Graph-based retrieval enhances logical consistency:} While GraphRAG alone scored lower in retrieval metrics, its structured approach ensures that responses maintain logical coherence, which is valuable for knowledge-intensive tasks.
    \item \textbf{Challenges in multi-document retrieval:} The lower performance in multi-paper retrieval underscores the need for improved retrieval ranking methods to mitigate irrelevant or redundant contexts.
\end{itemize}

To build on these findings, several avenues for future research are proposed:
\begin{itemize}
    \item \textbf{Enhanced Hybrid Retrieval Mechanisms:} Further optimizing the balance between vector-based and graph-based retrieval to improve efficiency and accuracy.
    \item \textbf{Adaptive Retrieval Strategies:} Implementing dynamic retrieval mechanisms that adjust based on query complexity and information density.
    \item \textbf{Expansion of Benchmarking Frameworks:} Introducing more comprehensive evaluation metrics, including human-in-the-loop assessments and real-world user testing.
    \item \textbf{Integration with Domain-Specific Knowledge:} Exploring the integration of domain-specific property graphs to enhance the retrieval precision in specialized fields.
\end{itemize}

This study provides a comparative analysis of different RAG approaches and highlights the advantages of combining vector and graph-based retrieval methods. The findings underscore the potential of Hybrid RAG to improve information retrieval accuracy, particularly for knowledge-intensive applications. Future research should focus on refining retrieval mechanisms and expanding evaluation frameworks to ensure the continued advancement of AI-driven literature review systems.

\section*{Acknowledgements}
The authors thank the Data Science and Automation Team at the Pest Management Regulatory Agency (PMRA) of Health Canada for their support and for providing the opportunity to conduct these experiments at their AI Lab. We also acknowledge Shared Services Canada for providing the cloud platform used to run the experiments. In addition, we thank the scientific evaluators at PMRA for their assistance in annotating and validating the test datasets.

\printbibliography[heading=subbibintoc]

@inproceedings{salemi2024evaluating,
  title={Evaluating retrieval quality in retrieval-augmented generation},
  author={Salemi, Alireza and Zamani, Hamed},
  booktitle={Proceedings of the 47th International ACM SIGIR Conference on Research and Development in Information Retrieval},
  pages={2395--2400},
  year={2024}
}

@article{lewis2020retrieval,
  title={Retrieval-augmented generation for knowledge-intensive nlp tasks},
  author={Lewis, Patrick and Perez, Ethan and Piktus, Aleksandra and Petroni, Fabio and Karpukhin, Vladimir and Goyal, Naman and K{\"u}ttler, Heinrich and Lewis, Mike and Yih, Wen-tau and Rockt{\"a}schel, Tim and others},
  journal={Advances in Neural Information Processing Systems},
  volume={33},
  pages={9459--9474},
  year={2020}
}

@article{gao2023retrieval,
  title={Retrieval-augmented generation for large language models: A survey},
  author={Gao, Yunfan and Xiong, Yun and Gao, Xinyu and Jia, Kangxiang and Pan, Jinliu and Bi, Yuxi and Dai, Yi and Sun, Jiawei and Wang, Haofen},
  journal={arXiv preprint arXiv:2312.10997},
  year={2023}
}

@article{jiang2024longrag,
  title={Longrag: Enhancing retrieval-augmented generation with long-context llms},
  author={Jiang, Ziyan and Ma, Xueguang and Chen, Wenhu},
  journal={arXiv preprint arXiv:2406.15319},
  year={2024}
}

@inproceedings{yu2024evaluation,
  title={Evaluation of retrieval-augmented generation: A survey},
  author={Yu, Hao and Gan, Aoran and Zhang, Kai and Tong, Shiwei and Liu, Qi and Liu, Zhaofeng},
  booktitle={CCF Conference on Big Data},
  pages={102--120},
  year={2024},
  organization={Springer}
}

@article{fan2025trustrag,
  title={TrustRAG: An Information Assistant with Retrieval Augmented Generation},
  author={Fan, Yixing and Yan, Qiang and Wang, Wenshan and Guo, Jiafeng and Zhang, Ruqing and Cheng, Xueqi},
  journal={arXiv preprint arXiv:2502.13719},
  year={2025}
}

@article{oro2025comprehensive,
  title={A Comprehensive Evaluation of Embedding Models and LLMs for IR and QA across English and Italian},
  author={Oro, Ermelinda},
  year={2025}
}

@article{chen2025pathrag,
  title={PathRAG: Pruning Graph-based Retrieval Augmented Generation with Relational Paths},
  author={Chen, Boyu and Guo, Zirui and Yang, Zidan and Chen, Yuluo and Chen, Junze and Liu, Zhenghao and Shi, Chuan and Yang, Cheng},
  journal={arXiv preprint arXiv:2502.14902},
  year={2025}
}

@article{ghanadian2024socially,
  title={Socially aware synthetic data generation for suicidal ideation detection using large language models},
  author={Ghanadian, Hamideh and Nejadgholi, Isar and Al Osman, Hussein},
  journal={IEEe Access},
  volume={12},
  pages={14350--14363},
  year={2024},
  publisher={IEEE}
}

@article{malin2024review,
  title={A review of faithfulness metrics for hallucination assessment in Large Language Models},
  author={Malin, B and Kalganova, T and Boulgouris, N},
  year={2024},
  publisher={Cornell University}
}

@misc{radford2021chatgptplus,
      title={Chat with GPT: Improving Language Generation and Task-Oriented Dialogue}, 
      author={Alec Radford and Ilya Sutskever and Rewon Child and Gretchen Krueger and Jong Wook Kim},
      year={2021},
      howpublished = {\url{https://openai.com/blog/chatgpt-plus}}
}

@misc{llamaindex_vectorstore,
  author = {{LlamaIndex Team}},
  title = {{VectorStoreIndex - LlamaIndex Documentation}},
  howpublished = {\url{https://docs.llamaindex.ai/en/stable/module_guides/indexing/vector_store_index/}},
  year = {2024},
  note = {Accessed: 2025}
}

@misc{llamaindex_propertygraph,
  author = {{LlamaIndex Team}},
  title = {{Property Graph Index - LlamaIndex Documentation}},
  howpublished = {\url{https://docs.llamaindex.ai/en/stable/module_guides/indexing/lpg_index_guide/}},
  year = {2024},
  note = {Accessed: 2025}
}

@misc{llamaindex_vectorretriever,
  author = {{LlamaIndex Team}},
  title = {{Vector Index Retriever - LlamaIndex API Reference}},
  howpublished = {\url{https://docs.llamaindex.ai/en/stable/api_reference/retrievers/vector/}},
  year = {2024},
  note = {Accessed: 2025}
}

@misc{llamaindex_vectorcontext,
  author = {{LlamaIndex Team}},
  title = {{Property Graph Index Retrievers - LlamaIndex Documentation}},
  howpublished = {\url{https://docs.llamaindex.ai/en/stable/module_guides/indexing/lpg_index_guide/}},
  year = {2024},
  note = {Accessed: 2025}
}

@misc{llamaindex_llmsynonym,
  author = {{LlamaIndex Team}},
  title = {{LLMSynonymRetriever - Property Graph Index Documentation}},
  howpublished = {\url{https://docs.llamaindex.ai/en/stable/module_guides/indexing/lpg_index_guide/}},
  year = {2024},
  note = {Accessed: 2025}
}

\end{document}